\documentclass{blois}

\def\be{\begin{equation}}
\def\ee{\end{equation}}
\def\bea{\begin{eqnarray}}
\def\eea{\end{eqnarray}}

\newcommand{\Photo}{\includegraphics[height=35mm]{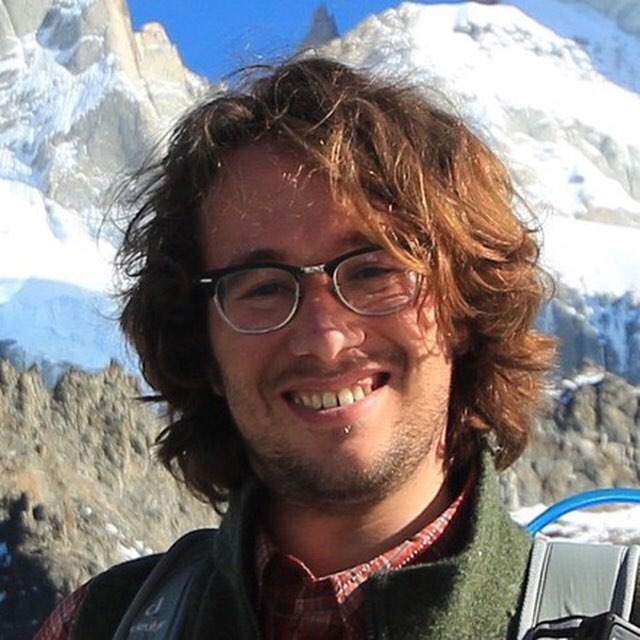}}

\begin{document}
\vspace*{4cm}
\title{$t\bar tH$ production in NNLO QCD}

\author{SIMONE DEVOTO\,\footnote{Work done in collaboration with Stefano Catani, Massimiliano Grazzini,
Stefan Kallweit, 
Javier Mazzitelli and Chiara Savoini.}}

\address{Dipartimento di Fisica Aldo Pontremoli, Universit\`a di Milano and INFN, Sezione di Milano, Via Celoria 16, I-20133 Milano, Italy}

\maketitle\abstracts{
       In this contribution we present our recent computation of the NNLO QCD corrections to the production of a Higgs boson associated with a top-antitop quark pair.
       This process is of great importance since it allows for a direct measurement of the top-quark Yukawa coupling, and the inclusion of NNLO corrections is crucial in order to provide theory predictions with an uncertainty competitive with the projected accuracy of the experimental measurements at the end of the high-luminosity phase of the LHC.
      }

\section{Introduction}
The discovery of the Higgs boson (H) in 2012~\cite{ATLAS:2012yve,CMS:2012qbp} confirmed one of the most glaring predictions of the Standard Model. Since then, studying its properties has been one of the priorities of the LHC scientific programme. In this context, its production associated with a pair of top quarks ($t\bar t$) plays a special role because of the strong coupling between the Higgs boson and the top quarks. Indeed, this same process allows for a direct measurement of the top quark Yukawa coupling.

The current theoretical uncertainties of the predictions for the $t\bar tH$ process are of ${\cal O}(10\%)$~\cite{LHCHiggsCrossSectionWorkingGroup:2016ypw}, and include the computation of next-to-leading order~(NLO) QCD corrections~\cite{Beenakker:2001rj,Beenakker:2002nc,Reina:2001sf,Reina:2001bc,Dawson:2002tg,Dawson:2003zu} and NLO EW corrections~\cite{Frixione:2014qaa,Yu:2014cka,Frixione:2015zaa}.

Beyond these on-shell calculations, the full off-shell process including leptonic top decays was studied at NLO QCD~\cite{Denner:2015yca,Stremmer:2021bnk} and NLO EW~\cite{Denner:2016wet}, and soft-gluon contributions close to the partonic kinematical threshold were resummed~\cite{Kulesza:2015vda,Broggio:2015lya,Broggio:2016lfj,Kulesza:2017ukk,Broggio:2019ewu,Kulesza:2020nfh}. 
As a first step towards next-to-next-to-leading order~(NNLO) QCD corrections, the contribution coming from off-diagonal channels has also been computed~\cite{Catani:2021cbl}.

On the experimental side, the $t\bar tH$ production signal has been measured by both the ATLAS and CMS collaborations with an uncertainty of ${\cal O}(20\%)$~\cite{Aaboud:2018urx,Sirunyan:2018hoz}, but the projection for the end of the High-Luminosity phase indicates an uncertainty at the ${\cal O}(2\%)$ level~\cite{Cepeda:2019klc}.

In this proceeding we report on our recent computation of the full NNLO QCD corrections for on-shell $t\bar t H$ production~\cite{Catani:2022mfv}, predictions that are required in order to match, on the theoretical side, the extraordinary precision expected for the future measurements. 
In order to estimate the unknown two-loop amplitudes for $t\bar t H$ production, we developed a soft-Higgs approximation, which will be discussed in the following.

\section{Challenges of the computation}
Our computation of the NNLO corrections to the $t\bar t H$ process
$c(p_1)+{\bar c}(p_2)\to t(p_3)+{\bar t}(p_4)+H(k)$ ($c=q,{\bar q},g$),
has been performed by using the $q_T$-subtraction formalism~\cite{Catani:2007vq} to handle and cancel singularities of infrared origin arising at intermediate steps of the computation.

Within the $q_T$-subtraction formalism, we can write the the cross section $d\sigma^{t\bar t H}$ as follows:
\begin{equation}
  \label{eq:qt_sub}
 d\sigma^{t\bar t H}_{NNLO}=\mathcal H^{t\bar t H}_{NNLO}\otimes d\sigma_{LO}^{t\bar t H}+\left[d\sigma_{NLO}^{t\bar t H+jet}-d\sigma^{CT}_{NNLO}\right]\;.
\end{equation}
The term in the square brackets in Eq.~(\ref{eq:qt_sub}) represents the contribution to the cross section with $q_T\neq 0$, $q_T$ being the transverse momentum of the final state system. As such, it is captured by the NLO cross section for the process $t\bar t H+$jet and can be computed with known NLO subtraction techniques\cite{Catani:1996vz}. The counterterm $d\sigma^{CT}_{NNLO}$ cancels additional singularities of pure NNLO type associated with the limit $q_T\to 0$. 

The contribution with exactly $q_T=0$ is provided by the coefficient $\mathcal H^{t\bar t H}_{NNLO}$:
\begin{equation}
\label{eq:H_factor}
    \mathcal H^{t\bar t H}_{NNLO}=H^{(2)}_{t\bar t H}\delta(1-z_1)\delta(1-z_2)+\delta\mathcal H^{(2)}_{t\bar t H}\;,
\end{equation}
which contains the main challenges that needed to be overcome in order to generalise the $q_T$-subtraction formalism for this process.

The factor $\delta\mathcal H^{(2)}_{t\bar t H}$ contains the known one-loop squared contribution and the soft parton contribution.
The latter has been computed in the case of heavy-quark pair production~\cite{Catani:2023tby} by assuming back-to-back kinematics for the massive quarks, and allowed us to apply $q_T$-subtraction to the case of top-pair production~\cite{Catani:2019iny,Catani:2019hip} and bottom-pair production~\cite{Catani:2020kkl}. In order to extend these results to the current class of processes, such a constraint had to be lifted: the extra contributions have been computed numerically and their on-the-fly evaluation has been implemented in a dedicated library~\cite{inprep}.

The coefficient $H^{(2)}_{t\bar t H}$ contains the genuine virtual contribution, in the form of the two-loop amplitude ${\cal M}^{(2)}_{t{\bar t}H}$: 
\begin{equation}
\label{eq:H2}
    H_{t\bar t H}^{(2)}=\frac{2\,Re\big(\mathcal M_{t\bar t H}^{(2)}(\mu_{IR},\mu_R)\mathcal M_{t\bar t H}^{(0)}\big)}{|\mathcal M_{t\bar t H}^{(0)}|^2}\;,
\end{equation}
where $\mu_{IR}$ is the scale at which the infrared poles are subtracted and $\mu_R$ is the renormalisation scale.
Since $\mathcal M_{t\bar t H}^{(2)}$ is not yet known, we had to estimate it via a suitable approximation.

In the limit in  which the momentum of the Higgs boson $k$ is soft ($k\to 0$), the amplitude $\mathcal M_{t\bar t H}(\{p_i\},k)$ fulfils:
\begin{equation}
\label{eq:fact}
\mathcal M_{t\bar t H}(\{p_i\},k)\simeq F(\alpha_S(\mu_R);m_t/\mu_R)\,\frac{m_t}{v}\sum_{i=3,4} \frac{m_t}{p_i \cdot k}\, \mathcal M_{t\bar t}(\{p_i\})\, ,
\end{equation}
where $v=(\sqrt{2}G_F)^{-1/2}$, and $\mathcal M_{t\bar t}(\{p_i\})$ is the amplitude in which the Higgs boson has been removed, i.e.\ the amplitude for $t\bar t$ production, available up to two-loop level~\cite{Barnreuther:2013qvf}.
The formula in Eq.~(\ref{eq:fact}) can be derived by using the eikonal approximation and low-energy theorems\cite{Kniehl:1995tn}. The function $F(\alpha_S(\mu_R);m_t/\mu_R)$ is the soft limit of the scalar form factor of the heavy-quark \cite{Bernreuther:2005gw,Ablinger:2017hst}.

In order to validate our approximation, we first test it at NLO, considering the contribution of the hard coefficient $H^{(1)}_{t\bar tH}$ to the NLO cross section and comparing the exact result with the approximated one. As it is shown in Table~\ref{tab:soft}, the deviation with respect to the exact computation is about $30\%$ for the $gg$ channel and $5\%$ for the $q\bar q$ channel.
The better agreement for the $q\bar q$ channel can be explained by the presence of additional diagrams where a Higgs boson is radiated from a virtual top, both at LO and NLO, in the $gg$ channel. Since the approximation captures the leading behaviour in the soft limit $k\to0$, the effect of the emission from highly off-shell top propagators as the ones in these families of diagrams is not correctly reproduced. 

\begin{table}
\begin{center}
\begin{tabular}{|c|ll|ll|}\hline
& \multicolumn{2}{c|}{$\sqrt{s}=13\,\mathrm{TeV}$} & \multicolumn{2}{c|}{$\sqrt{s}=100\,\mathrm{TeV}$}\\
& \multicolumn{1}{c}{$gg$} & \multicolumn{1}{c|}{$q{\bar q}$} & \multicolumn{1}{c}{$gg$} & \multicolumn{1}{c|}{$q{\bar q}$}\\
\hline
$\Delta\sigma_{\rm NLO,H}$ [fb]& $\phantom{-0}88.62$  & $\phantom{00}7.826$ & $\phantom{-0}8205$ & $\phantom{0}217.0$\\
$\Delta\sigma_{\rm NLO,H}|_{\rm soft}$ [fb]& $\phantom{-0}61.98$ & $\phantom{00}7.413$ & $\phantom{-0}5612$ & $\phantom{0}206.0$\\
\hline
Difference & $\phantom{00}30.1\%$ & $\phantom{00}5.27\%$ & $\phantom{00}31.6\%$ & $\phantom{00}5.06\%$\\
\hline
\end{tabular}
\end{center}
\caption{\label{tab:soft}Hard contribution to the NLO cross sections in the soft approximation.}
\end{table}

Having checked the behaviour of the approximation at NLO, we can employ it for the computation of the NNLO corrections.
We find a small contribution from the (approximated) NNLO hard coefficient $H^{(2)}_{t\bar tH}$: it corresponds to $1\%$ of the LO cross-section in the $gg$ channel and to $2\%$ of the LO cross-section in the $q\bar q$ channel.
To estimate the uncertainties due to the approximation at NNLO, we use the results in Table~\ref{tab:soft}:
we consider the deviation from the exact results at NLO as a lower bound on the NNLO uncertainty and we multiply it by a tolerance factor of 3, finally combining linearly the uncertainties for the $q\bar q$ and $gg$ channel.
With this choice, the final uncertainties on the NNLO corrections amount to $\pm15\%$, which corresponds to $\pm0.6\%$ on the total cross section. 

In order to test if such estimation of the uncertainties is reliable, we can compare it with the effect given by the variation of (unphysical) parameters of the approximation. 
A first check that can be performed is to use different recoil prescriptions.
In order to apply the approximation, there is the need to map the $2\to3$ kinematics of $t\bar t H$ production to the $2\to 2$ kinematics of $t\bar t$ production. 
In our computation, we use the $q_T$ recoil prescription~\cite{Catani:2015vma}, reabsorbing the Higgs momentum equally in the initial-state parton momenta and leaving unchanged the top and anti-top momenta.
We verified that the difference in the result obtained by using different prescriptions, for example by reabsorbing the transverse momentum of the Higgs boson entirely into one of the initial state momenta, is negligible when compared with the provided uncertainties.
Another possible check is to vary the scale $\mu_{IR}$ at which the infrared singularities are subtracted, which in our computation we fix at the virtuality of the $t\bar t H$ system $M$.
In an exact computation, the dependence on this parameter would cancel between $H^{(2)}_{t\bar t H}$ and $\delta\mathcal H^{(2)}_{t\bar t H}$ in Eq.~(\ref{eq:H_factor}).
By repeating our calculation with $\mu_{IR}=2M$ and $\mu_{IR}=M/2$ we verified that the numerical result for the hard virtual contribution changes by ${}^{+164\%}_{-25\%}$ (${}^{+142\%}_{-20\%}$) in the $gg$ channel and by ${}^{+4\%}_{-0\%}$ (${}^{+3\%}_{-0\%}$) in the $q{\bar q}$ channel at \mbox{$\sqrt{s}=$ 13(100)\,TeV}, an amount comparable with the uncertainties that have been estimated.

\section{Results}

Our results for the inclusive cross section, obtained by implementing the $t\bar t H$ process within the {\sc Matrix} framework~\cite{Grazzini:2017mhc}, are shown in Table~\ref{tab:res} (see Ref.~\cite{Catani:2022mfv}), together with the customary 7-point scale variation. 
For the NNLO predictions, we quote in brackets the combination of the numerical errors and the soft-approximation uncertainty described in the previous section: it can be observed that they turn out to be significantly smaller than remaining perturbative uncertainties.

The inclusion of NNLO corrections leads to an increase of $+4\%$ at centre-of-mass energy $\sqrt{s}=13$ TeV and of $+2\%$ at $\sqrt{s}=100$ TeV. In both cases, they also significantly reduce the scale-variation bands, which are at the few percent level. 

\begin{figure}
\centerline{\includegraphics[width=0.5\linewidth]{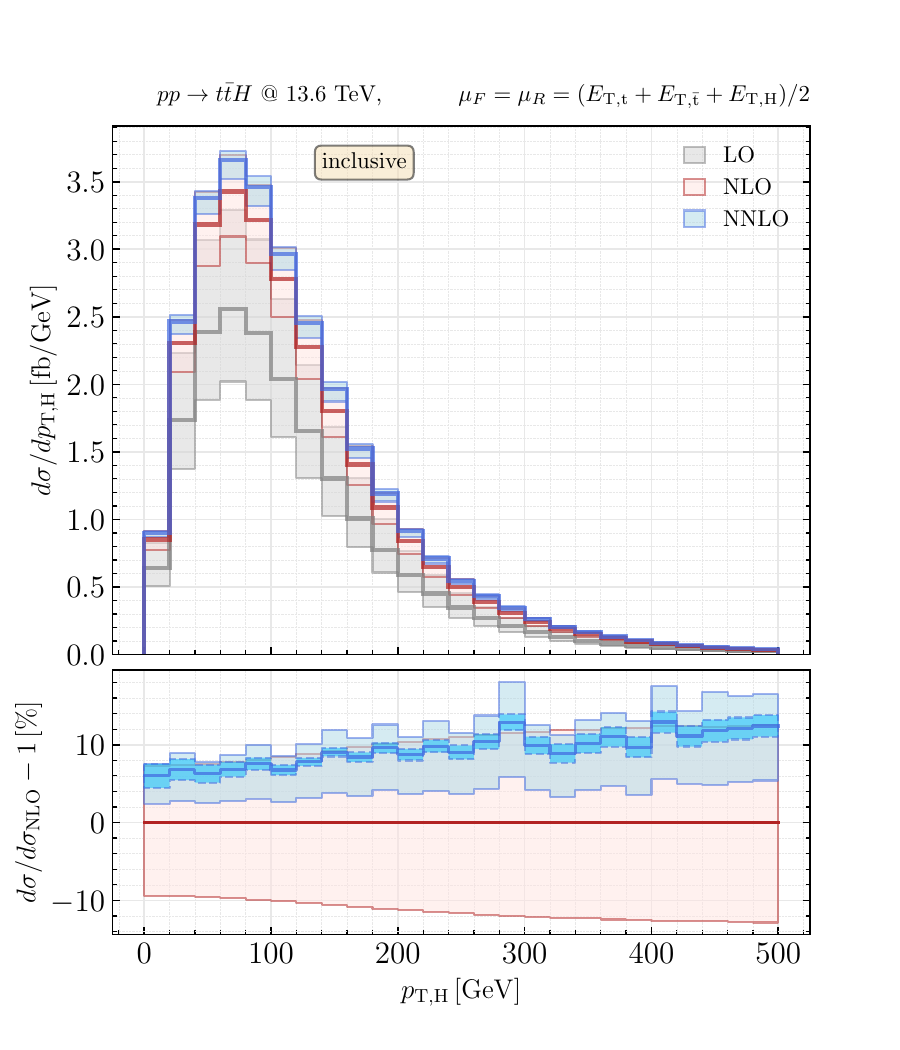}}
\caption[]{Single-differential cross sections as a function of $p_{T,H}$ ({\sc preliminary}).}
\label{fig:diff}
\end{figure}

In Figure~\ref{fig:diff} we also present some preliminary results for the transverse-momentum distribution of the Higgs boson.
In the upper plot we show our LO (grey), NLO (red) and NNLO (blue) predictions, with their scale variation bands, while in the lower plot the ratio to the NLO prediction is presented. 
The smaller dark blue band in the NNLO predictions shows the soft-approximation uncertainties, computed as described above and on a bin-by-bin basis.
We can observe that the NLO and NNLO uncertainty bands overlap, providing first signs of perturbative convergence.
Despite the fact that the soft approximation is expected to be less accurate at high $p_{T,H}$ values, the uncertainties remain of the same order over all the spectrum. 
This can be understood from the fact that at high $p_{T,H}$ the role of the $gg$ channel is reduced and the $q\bar q$ channel, which is under better control, plays the major role.

\begin{table}[t]
\begin{center}
\begin{tabular}{|c|l|l|}\hline
$\sigma$ [pb] & \multicolumn{1}{c|}{$\sqrt{s}=13\,\mathrm{TeV}$} & \multicolumn{1}{c|}{$\sqrt{s}=100\,\mathrm{TeV}$}\\
\hline
$\sigma_{\rm LO}$ & $0.3910\,^{+31.3\%}_{-22.2\%}$ & $25.38\,^{+21.1\%}_{-16.0\%}$\\
$\sigma_{\rm NLO}$ & $0.4875\,^{+5.6\%}_{-9.1\%}$ & $36.43\,^{+9.4\%}_{-8.7\%}$\\
$\sigma_{\rm NNLO}$ & $0.5070\,(31)^{+0.9\%}_{-3.0\%}$ & $37.20(25)\,^{+0.1\%}_{-2.2\%}$\\
\hline
\end{tabular}
\end{center}
\caption{
\label{tab:res}
LO, NLO and NNLO cross sections at $\sqrt{s}=13$\,TeV and $\sqrt{s}=100$\,TeV. The errors stated in brackets at NNLO combine numerical errors with the uncertainty due to the soft Higgs boson approximation.
}
\end{table}

\section{Conclusions}
In this contribution we reported on our recent computation of the NNLO QCD corrections to $t\bar tH$ production at hadron colliders. Our computation is exact, with the exception of the contribution from the two-loop amplitudes which have been estimated by using a soft-Higgs approximation.
The uncertainty due to such approximation is estimated to be smaller than $1\%$ of the inclusive result, and well below the remaining perturbative uncertainties.
The inclusion of NNLO QCD corrections lead to an increase of the cross section of a few percent, and to a significant reduction of the scale-variation bands.
We have also presented preliminary results for the transverse momentum distribution of the Higgs boson.

\section*{References}

\bibliography{biblio}

\end{document}